\documentclass[aps, showpacs, twocolumn, preprintnumbers, 
nofootinbib, nobibnotes, superscriptaddress, longbibliography, 
amsmath, amssymb, prd, 10pt, floatfix]{revtex4-2}

\usepackage[skins,theorems]{tcolorbox}
\usepackage{color}
\usepackage{soul}
\usepackage{centernot}
\usepackage{amsfonts} 
\usepackage{graphicx} 
\graphicspath{{./figures/}} 
\usepackage{dcolumn} 
\usepackage{siunitx}
\usepackage{bm} 
\usepackage{bbold}
\usepackage{braket}
\usepackage{esdiff} 
\usepackage[mathlines]{lineno} 
\usepackage{subfigure}
\usepackage{parskip} 
\usepackage{dsfont}
\usepackage{float} 
\usepackage{url} 
\usepackage{xcolor}
\usepackage[
    colorlinks=true,
    linkcolor=blue,
    citecolor=blue,
    urlcolor=blue,
    pdftitle={T\'itulo del Documento},
    pdfauthor={Tu Nombre},
    unicode=true
]{hyperref} 

\hypersetup{
    colorlinks=true,       
    citecolor=red,         
    linkcolor=green,       
    urlcolor=blue          
}

\showboxbreadth\maxdimen\showboxdepth\maxdimen

\begin{document}
\title{Economic Entropy and Sectoral Dynamics: A Thermodynamic Approach to Market Analysis }

\author{W. A. Rojas C.}
\email{warojasc@unal.edu.co}
\affiliation{Universidad Distrital Francisco Jos\'e de Caldas, Bogot\'a, Colombia}

\author{A. Zamora V.}
\email{azamorav@udistrital.edu.co}
\affiliation{Universidad Distrital Francisco Jos\'e de Caldas, Bogot\'a, Colombia}

\author{L. F. Quijano W.}
\email{lfquijanow@udistrital.edu.co}
\affiliation{Universidad Distrital Francisco Jos\'e de Caldas, Bogot\'a, Colombia}

\author{Y. Beltran P.}
\email{nybeltranp@udistrital.edu.co}
\affiliation{Universidad Distrital Francisco Jos\'e de Caldas, Bogot\'a, Colombia}


\begin{abstract} 

This paper presents an  application of geometrothermodynamics (GTD) to the economic analysis of Bogot\'a's sports sector through the Satellite Account of Sport (CSDB). By establishing an analogy between thermodynamic systems and economic structures, we develop a mathematical framework where monetary flows behave analogously to energy, while economic entropy, temperature, and heat capacity acquire well-defined economic interpretations. The study focuses on two contrasting sectors: gambling and betting $\mathbb{S}_{15}$, and recreational and sports activities $\mathbb{S}_{16}$, analyzing data from 2018-2023. Our results demonstrate that $\mathbb{S}_{15}$ exhibits lower economic entropy than  $\mathbb{S}_{16}$ , indicating a higher degree of organization and regulatory structure in the gambling sector compared to the more heterogeneous recreational sports sector. The heat capacity function reveals critical points that may signal phase transitions in economic dynamics, while Ricci and Kretschmann curvature scalars identify potential crisis points in the sectoral organization. Furthermore, the cross-income elasticity analysis shows distinct resource flow patterns between sectors, suggesting that gambling activities may serve as an economic driver for recreational sports. This thermodynamic approach provides a quantitative tool for analyzing resource redistribution policies and anticipating critical transitions in sectoral economics. The findings suggest that econophysics and statistical thermodynamics constitute powerful frameworks for understanding the sectoral dynamics of Bogot\'a's sports economy, with significant potential for developing prospective analysis tools in public policy design.
\end{abstract} 
\maketitle

\section{Introduction}\label{sec1}

The economy can be understood as a complex system in which agents—such as consumers and producers—continuously interact to make decisions regarding the production, distribution, and consumption of goods and services. Among the microeconomic factors guiding these interactions, supply and demand stand out; their dynamics determine equilibrium prices and quantities. Likewise, consumer preferences shape market structures and influence resource allocation \cite{farmer2009, mas1995,marshall1890}. 

In the late 1990s, a research field known as  econophysics  emerged, proposing the use of tools from statistical mechanics to analyze income and wealth distributions. One of its most significant findings is the observation of a bimodal structure: the majority of the population follows a Boltzmann–Gibbs–type exponential distribution, whereas a small minority is described by a power-law tail, associated with Pareto’s law \cite{DragulescuYakovenko2001,DragulescuYakovenko2003,dragulescu2000statistical,draganulescu2001evidence}.
This duality has been empirically confirmed across various countries, with the concept of  economic temperature  serving as a parameter that enables comparisons of average income across economies and provides an analogy for wealth flows akin to heat flow.

Subsequently, kinetic models were developed in which economic agents behave like particles in a gas, exchanging money under conservation principles. These approaches demonstrate that introducing a propensity to save yields more realistic stationary distributions, reproducing the coexistence of an exponential component and a Pareto tail. Such results allow the reinterpretation of economic utility maximization as equivalent to entropy maximization, thereby strengthening the formal analogy between microeconomics and thermodynamics.\cite{chakraborti2000statistical,Chakrabarti2013}. 
 
Comprehensive reviews, such as those by Yakovenko \cite{Yakovenko2008,yakovenko2009colloquium} consolidated this framework by demonstrating that both empirical data and computational simulations confirm the robustness of the model. In parallel, alternative interpretations were proposed: Mimkes \cite{Mimkes2006} introduced two economic laws inspired by classical thermodynamics, defining an economic entropy associated with the distribution of goods and income, while Rawlings \cite{RAWLINGS2004643} explored the conservation of logarithms of income as a basis for the emergence of Pareto’s law.

More recently, authors such as Costa \cite{Costa2024} and Lozada \cite{lozada2017hotelling} have extended these analogies to broader frameworks, introducing explicit formulations of economic laws inspired by thermodynamics and discussing the role of entropy in the sustainability of economic growth. Meanwhile, Quevedo and collaborators \cite{Quevedo2011,Quevedo2023} have formalized Geometrothermodynamics (GTD) as a geometric framework that associates Legendre-invariant metrics with economic systems, providing a tool to analyze phase transitions and financial crises.

Finally, recent works such as \cite{Rojas_Zamora_2025} apply these ideas to specific contexts, such as Bogotá’s sports sector. There, the combined use of statistical mechanics and GTD enables the characterization of income elasticities, the computation of thermodynamic potentials (e.g., entropy, heat capacity), and the detection of singularities linked to periods of economic instability—particularly during the COVID-19 pandemic.

\section{Elements of Statistical Thermodynamics Applied to Economic Systems}\label{sec3}
Quevedo et al. \cite{Quevedo2011,Quevedo2016,Quevedo2023} propose a statistical–physical model to describe economic systems, in which the conservation of a total amount of money $M$ over a time interval $t$ is postulated, expressed as
\begin{equation}
    \frac{dM}{dt} = 0.
    \label{eqn00}
\end{equation}

Let  $N$ denote the number of economic agents sharing this total amount of money \cite{Santos_2016}. Then, $M$ can be decomposed as the sum of the individual monetary amounts $m_{i}$ held by each agent:
\begin{align}
    M &= m_{1} + m_{2} + \cdots + m_{N} \notag \\
      &= \sum_{i=1}^{N} m_{i}.
    \label{eqn10}
\end{align}

Within the canonical ensemble framework, the probability distribution of money among agents follows an exponential law:
\begin{equation}
    \rho(m) \propto e^{-m/T},
    \label{eqn20}
\end{equation}
where $T = \langle m \rangle$ represents the average amount of money per agent (analogous to temperature in thermodynamics), and $m = m(\bar{\lambda})$  is the function describing the amount of money accessible to an agent in terms of a set of microeconomic parameters $\bar{\lambda} = (\lambda_1, \lambda_2, \ldots)$.

The corresponding normalized distribution is:
\begin{equation}
    \rho(\bar{\lambda}) = \frac{e^{-m(\bar{\lambda})/T}}{Z(T, \bar{x})},
    \label{eqn30}
\end{equation}
where $Z(T, \bar{x})$  is the partition function of the system, defined as:

\begin{equation}
    Z(T, \bar{x}) = \int e^{-m(\bar{\lambda})/T} \, d\bar{\lambda},
    \label{eqn40}
\end{equation}
with $\bar{x}$  denoting a set of macroeconomic variables that characterize the state of the system.

The expected value of any observable $g=g(\bar{\lambda})$ is computed as:

\begin{align}
    \langle g \rangle &= \int g(\bar{\lambda}) \, \rho(\bar{\lambda}) \, d\bar{\lambda} \notag \\
                     &= \frac{1}{Z(T, \bar{x})} \int g(\bar{\lambda}) \, e^{-m(\bar{\lambda})/T} \, d\bar{\lambda}.
    \label{eqn50}
\end{align}

In particular, the average money per agent is:
\begin{equation}
    \langle m \rangle = \int m(\bar{\lambda}) \, \rho(\bar{\lambda}) \, d\bar{\lambda}.
    \label{eqn60}
\end{equation}
Differentiating this expression yields:
\begin{equation}
    d\langle m \rangle = \int m \, d\rho \, d\bar{\lambda} + \int \rho \, dm \, d\bar{\lambda}.
    \label{eqn70}
\end{equation}

Defining the term:
\begin{equation}
    \langle dm \rangle = \int \rho \, dm \, d\bar{\lambda},
    \label{eqn80}
\end{equation}
Equation (8) is then rewritten as
\begin{equation}
    d\langle m \rangle = \int m \, d\rho \, d\bar{\lambda} + \langle dm \rangle.
    \label{eqn90}
\end{equation}
From  \eqref{eqn30}, we can solve for $m(\bar{\lambda})$:
\begin{equation}
    m(\bar{\lambda}) = -T \left[ \ln|\rho(\bar{\lambda})| + \ln|Z(T, \bar{x})| \right].
    \label{eqn100}
\end{equation}
Since  $\rho(\bar{\lambda})$ is a normalized probability distribution:
\begin{equation}
    \int \rho(\bar{\lambda}) \, d\bar{\lambda} = 1, \quad d\int \rho(\bar{\lambda}) \, d\bar{\lambda} = 0,
    \label{eqn110}
\end{equation}
the second term in the differential of $m$ vanishes upon integration.

Substituting \eqref{eqn100} into the first term of \eqref{eqn70} yields:

\begin{align}
    \int m \, d\rho \, d\bar{\lambda} &= -T \int \left[ \ln|\rho(\bar{\lambda})| + \ln|Z(T, \bar{x})| \right] d\rho \, d\bar{\lambda} \notag \\
                                      &= -T \int \ln|\rho(\bar{\lambda})| \, d\rho \, d\bar{\lambda},
    \label{eqn120}
\end{align}
where the term involving \(\ln|Z|\) vanishes by \eqref{eqn110}.

The entropy of the economic system is then defined as:

\begin{align}
    S &= \langle -\ln|\rho(\bar{\lambda})| \rangle \notag \\
      &= -\int \rho(\bar{\lambda}) \ln|\rho(\bar{\lambda})| \, d\bar{\lambda},
    \label{eqn130}
\end{align}
whose differential is:
\begin{equation}
    dS = -\int \ln|\rho(\bar{\lambda})| \, d\rho(\bar{\lambda}) \, d\bar{\lambda}.
    \label{eqn140}
\end{equation}
On the other hand, consider the derivative of money with respect to a macroeconomic parameter $x_{i}$:

\begin{align}
    y_i &= \left\langle -\frac{\partial m}{\partial x_i} \right\rangle \notag \\
        &= -\int \frac{\partial m}{\partial x_i} \, \rho(\bar{\lambda}) \, d\bar{\lambda}.
    \label{eqn150}
\end{align}
Then, the variation of money associated with changes in $x_{i}$ is:

\begin{equation}
    -y_i \, dx_i = \int \rho(\bar{\lambda}) \, dm \, d\bar{\lambda},
    \label{eqn160}
\end{equation}
and, in general:
\begin{equation}
    \langle dm \rangle = -\sum_{i=1}^{N} y_i \, dx_i.
    \label{eqn170}
\end{equation}

Substituting \eqref{eqn140} and \eqref{eqn170} into \eqref{eqn90} yields the First Law of Thermodynamics for economic systems \cite{Quevedo2011,RAWLINGS2004643}
\begin{equation}
    d\langle m \rangle = T \, dS - \sum_{i=1}^{N} y_i \, dx_i,
    \label{eqn180}
\end{equation}

where heat is identified as $dQ = T\,dS$ and economic work as $dW = \sum_i y_i \, dx_i$.

From \eqref{eqn100}, we can explicitly compute the entropy. Substituting $\ln|\rho(\bar{\lambda})|$:
\begin{equation}
    \ln|\rho(\bar{\lambda})| = -\frac{m(\bar{\lambda})}{T} - \ln|Z(T, \bar{x})|,
    \label{eqn190}
\end{equation}
so
\begin{equation}
    S = \int \left( \frac{m(\bar{\lambda})}{T} + \ln|Z(T, \bar{x})| \right) \rho(\bar{\lambda}) \, d\bar{\lambda}.
    \label{eqn200}
\end{equation}

Analogously to the Helmholtz free energy in thermodynamics, the free money function is defined as:

\begin{equation}
    f = \langle m \rangle - T S,
    \label{eqn210}
\end{equation}
which can be expressed in terms of the partition function as
\begin{equation}
    f = -T \ln|Z(T, \bar{x})|.
    \label{eqn220}
\end{equation}
From this function, key thermodynamic relations are derived:

\begin{equation}
    S = -\left( \frac{\partial f}{\partial T} \right)_{\bar{x}},
    \label{eqn230}
\end{equation}
\begin{equation}
    y_i = -\left( \frac{\partial f}{\partial x_i} \right)_{T, \{x_{j \neq i}\}},
    \label{eqn240}
\end{equation}
where partial derivatives are taken while holding all other macroscopic parameters constant.  
The heat capacity of the economic system is defined as:
\begin{equation}
    C = T \frac{\partial S}{\partial T},
    \label{eqn241}
\end{equation}
and provides a measure of the system’s response to changes in the economic temperature $T$. 

From Eq. (26), it is possible to estimate the heat produced by the sector, following the analogy with classical thermodynamics [21]:
\begin{equation}
C=\frac{dQ}{dT}
\label{eqn242}
\end{equation}
\section{Geometrothermodynamics for Economic Systems}
Quevedo \textit{et al.}  argue that an economic system, in addition to being a thermodynamic system—i.e., possessing a temperature $T$, an entropy $S$, and other properties characteristic of physical systems—can be described within a geometric framework that captures its intrinsic thermodynamic structure \cite{Quevedo2011}.

Geometrothermodynamics consists in introducing a metric on the equilibrium space $\mathcal{E}$ , such that points ($P\in\mathcal{E}$) represent all possible equilibrium states of the system \cite{Quevedo2023,Quevedo2007,Larranaga2011,valdes2016interpretacion,pineda2019geometrotermodinamica}. This endows $\mathcal{E}$ with a Riemannian geometric structure characterized by a specific metric tensor, from which one can compute curvature tensors such as the Riemann tensor $R_{abcd}$, $R_{ab}$, $K^{abcd}_{\,\,\,abcd}$ and $R^{a}_{\,\,\,a}$.

Let $\Phi(E^{a})$ , with $a=1,\ldots n$ , denote the fundamental equation of the system, where $\Phi$ is the thermodynamic potential, $E^{a}$ are the extensive variables—which serve as coordinates on the equilibrium space $ \mathcal{E}$—and $n$ is the number of degrees of freedom. The Hessian metric is then given by
\begin{equation}
g^{H}=\frac{\partial^{2}\Phi}{\partial E^{a}\partial E^{B}}dE^{a}dE^{b}.
\label{gtd00}
\end{equation}
Consequently, the equilibrium space is regarded as

\begin{equation}
\Phi =
\left\{ \begin{array}{rcl}
U\longrightarrow &  g^{W}\,\,   \mbox{Weinhold metric}
\\ 
-S\longrightarrow & g^{R}\,\,   \mbox{Ruppeiner metric}. 
\end{array}\right\}
\label{gtd10}
\end{equation}

The thermodynamic description of a physical system must be independent of the choice of thermodynamic potential—a property known as Legendre invariance. Accordingly, one defines metrics $g$ on $\mathcal{E}$ that satisfy this invariance principle. A drawback of \eqref{gtd00} is that it does not obey Legendre invariance. Nevertheless, Geometrothermodynamics (GTD) provides a family of metrics that fulfill this requirement.

\begin{equation}
g^{I}=\beta_{\Phi}\Phi\delta^{c}_{a}\frac{\partial^{2}\Phi}{\partial E^{b}\partial E^{c}},
\label{gtd20}
\end{equation}
\begin{equation}
g^{II}=\beta_{\Phi}\Phi\eta^{c}_{a}\frac{\partial^{2}\Phi}{\partial E^{b}\partial E^{c}},
\label{gtd30}
\end{equation}
\begin{equation}
g^{III}=\sum^{n}_{a=1}    \left[ \delta_{ab}E^{d}\frac{\partial \Phi}{\partial E^{a}}\right]\delta^{ab}\frac{\partial^{2}\Phi}{\partial E^{b}\partial E^{c}}dE^{a}dE^{c},
\label{gtd40}
\end{equation}
where $\delta^{c}_{a}=\mbox{diag}(1,\ldots ,1)$, $\eta^{c}_{a}=\mbox{diag}(-1,1,\ldots ,1)$ , and $\beta_{\Phi}$ denotes the degree of homogeneity of the thermodynamic potential $\Phi$ \cite{valdes2016interpretacion,pineda2019geometrotermodinamica}.


\section{Sports economic system}
The Sports Satellite Account (SSA-Bogotá, or CSDB) \cite{dane2023} is a statistical framework that links sports-related activities with complementary productive sectors. It is jointly developed by Colombia’s National Administrative Department of Statistics (DANE) \cite{dane2023} and Bogotá’s District Institute of Recreation and Sport (IDRD) \cite{idrd2023}. The CSDB encompasses 17 economic sectors that contribute to the sports economy, ranging from manufacturing and production activities (\(\mathbb{S}_{1}\)–\(\mathbb{S}_{7}\)), construction (\(\mathbb{S}_{8}\)), commerce (\(\mathbb{S}_{9}\)), programming, broadcasting, and transmission services (\(\mathbb{S}_{10}\)), leasing and rental services (\(\mathbb{S}_{11}\)), public administration and defense (\(\mathbb{S}_{12}\)), education (\(\mathbb{S}_{13}\)), human health (\(\mathbb{S}_{14}\)), gambling and betting (\(\mathbb{S}_{15}\)), sports, recreational, and entertainment activities (\(\mathbb{S}_{16}\)), to membership organizations (\(\mathbb{S}_{17}\)) \cite{dane2023}. A preliminary review of the CSDB for the period 2018–2023 allows identification of the sectors \(\mathbb{S}_{i} \subset \text{CSDB}\) that contribute most significantly (see Table~\ref{Tab01}).

\begin{table}[htbp]
\centering
\begin{tabular}{| c | c |}
\hline
$\left\langle \mathbb{S}_{i}\right\rangle$ & COP \,\,\,$*10^{11}$\\ \hline
$\left\langle \mathbb{S}_{15}\right\rangle$ & 1.949 \\
$\left\langle \mathbb{S}_{9}\right\rangle$  & 1.680 \\
$\left\langle \mathbb{S}_{16}\right\rangle$ & 1.158 \\
$\left\langle \mathbb{S}_{12}\right\rangle$ & 1.008 \\
$\left\langle \mathbb{S}_{2}\right\rangle$  & 0.427 \\ \hline
\end{tabular}
\caption{Average value of the sectors \(\mathbb{S}_{i}\) that contribute most to the CSDB over the period 2018–2023.}
\label{Tab01}
\end{table}

We analyze the cross-income elasticity \(\lambda_{\mathbb{S}_{i}}\) \cite{Rojas_Zamora_2025}.

\begin{equation}
\lambda_{\mathbb{S}_{i}}=\frac{\frac{\Delta \mathbb{S}_{i}}{\left\langle \mathbb{S}_{i}\right\rangle}}{\frac{\Delta CSDB}{\left\langle CSDB\right\rangle}}
\label{aqn000}
\end{equation}

\begin{equation}
\Delta CSDB=CSDB_{i+1}-CSDB_{i},
\label{aqn010}
\end{equation}

	\[\left\langle CSDB\right\rangle=\frac{CSDB_{i+1}+CSDB_{i}}{2},
\]

\begin{equation}
\Delta \mathbb{S}_{i}=\mathbb{S}_{i+1}-\mathbb{S}_{i},\,\,\,\left\langle \mathbb{S}_{i}\right\rangle=\frac{\mathbb{S}_{i+1}+\mathbb{S}_{i}}{2}
\label{aqn020}
\end{equation}
for the sectors \(\mathbb{S}_{15}\) and \(\mathbb{S}_{16}\), which have the greatest impact on the CSDB. In this context, it is useful to consider the contribution of both microeconomic and macroeconomic variables relevant to the CSDB that enter into the construction of the partition function, as follows:

\begin{table}[htbp]
\centering
\begin{tabular}{| c | c | c | c |}
\hline
Parameter & Symbol & Dimension & Type \\ \hline
 & & & \\
Money function & $m\left(\bar{\lambda}\right)$ & $\left[\frac{D}{E}\right]$ & Microeconomic \\
Elasticity & $\lambda_{\mathbb{S}_{i}}$ & $\left[1\right]$ & Microeconomic \\
Producer Price Index (IPP) & $\pi_{\mbox{\tiny{IPP}}}$ & $\left[1\right]$ & Macroeconomic \\
Consumer Price Index (IPC) & $\pi_{\mbox{\tiny{IPC}}}$ & $\left[1\right]$ & Macroeconomic \\
Exchange Rate (COP/USD) & $\sigma_{\mbox{\tiny{TRM}}}$ & $\left[\frac{D}{\text{Div}}\right]$ & Macroeconomic \\
 & & & \\
\hline
\end{tabular}
\caption{Description of CSDB parameters considered over the period 2018--2023.}
\label{Tab010}
\end{table}

where \( m\left(\bar{\lambda}\right) \) is the money function defined in \eqref{eqn40}, \( \lambda_{\mathbb{S}_{i}} \) denotes the elasticity coefficient, \( \pi_{\mbox{\tiny{IPP}}} \) is the producer price index (IPP) function, \( \pi_{\mbox{\tiny{IPC}}} \) is the consumer price index (IPC) function, and \( \sigma_{\mbox{\tiny{TRM}}} \) represents the market representative exchange rate (TRM) function. Normalization of \( \sigma_{\mbox{\tiny{TRM}}} \) allows the definition of a dimensionless variable \cite{daneipp,daneipc,banreptrm}.
\begin{equation}
\lambda_{\mbox{\tiny{TRM}}}=\frac{\sigma_{\mbox{\tiny{TRM},t}}}{\sigma_{\mbox{\tiny{TRM},0}}},
\label{aqn030}
\end{equation}
where \(\sigma_{\mbox{\tiny{TRM},t}}\) denotes the TRM value at time \(t\), and \(\sigma_{\mbox{\tiny{TRM},0}}\) is the TRM at \(t = 2018\). This enables us to identify the money function in \eqref{eqn40} as

\begin{align}
    m_{\mathbb{S}_{i}}(\bar{\lambda},\bar{\Lambda})	&=k_{\mathbb{S}_{i}}f(\lambda_{\mathbb{S}_{i}},\pi_{\mbox{\tiny{IPP}}},\pi_{\mbox{\tiny{IPC}}}, \lambda_{\mbox{\tiny{TRM}}}) 
\notag\\
					&=k_{\mathbb{S}_{i}}\lambda^{v_{\mathbb{S}_{i}}}_{\mathbb{S}_{i}}\pi_{\mbox{\tiny{IPP}}} ^{x_{\mathbb{S}_{i}}}\pi_{\mbox{\tiny{IPC}}}^{y_{\mathbb{S}_{i}}}\lambda_{\mbox{\tiny{TRM}}}^{z_{\mathbb{S}_{i}}}
&\hspace{0.3cm}
\label{aqn040}
\end{align}
where we assume a power–law relationship and statistical independence among the microeconomic and macroeconomic parameters listed in Table \ref{Tab010}. Moreover, for \eqref{aqn040}, the exponents \( v_{\mathbb{S}_{i}} \), \( x_{\mathbb{S}_{i}} \), \( y_{\mathbb{S}_{i}} \), and \( z_{\mathbb{S}_{i}} \) are determined through multiple logarithmic regression analysis.

\begin{widetext} 
\begin{align}
    \ln \left|m_{\mathbb{S}_{i}}(\bar{\lambda},\bar{\Lambda})	\right|&= \ln \left|k_{\mathbb{S}_{i}}\lambda^{v_{\mathbb{S}_{i}}}_{\mathbb{S}_{i}}\pi_{\mbox{\tiny{IPP}}} ^{x_{\mathbb{S}_{i}}}\pi_{\mbox{\tiny{IPC}}}^{y_{\mathbb{S}_{i}}}\lambda^{z_{\mathbb{S}_{i}}}_{\mbox{\tiny{TRM}}}\right|
\notag\\
		&= \ln \left|k_{\mathbb{S}_{i}}\right|+ \ln \left| \lambda^{v_{\mathbb{S}_{i}}}_{\mathbb{S}_{i}}\right|+ \ln \left|\pi_{\mbox{\tiny{IPP}}} ^{x_{\mathbb{S}_{i}}}\right|+ \ln\left|\pi_{\mbox{\tiny{IPC}}}^{y_{\mathbb{S}_{i}}}\right|+ \ln \left|\lambda_{\mbox{\tiny{TRM}}}^{z_{\mathbb{S}_{i}}}\right|
		\notag\\
		Y&=b_{0}+b_{1}X_{1}+b_{2}X_{2}+b_{3}X_{3}+b_{4}X_{4},
&\hspace{0.3cm}
\label{aqn050}
\end{align}
\end{widetext}
From \eqref{aqn060}, one can define the design matrix \cite{timm2007applied}:
\begin{equation}
\beta=\left(X^{T}X\right)^{-1}X^{T}Y.
\label{aqn060}
\end{equation}
The data are normalized when
\begin{equation}
X_{N}=\frac{X-\mu_{X}}{\sigma_{X}},
\label{aqn070}
\end{equation}
where \(\mu_{X}\) is the mean and \(\sigma_{X}\) is the standard deviation. Thus, for the CSDB statistics \cite{dane2023}, one can define
\begin{equation}
\left|\beta\right|=\begin{pmatrix}
\ln  k_{
\mathbb{S}_{i}}\\
v_{\mathbb{S}_{i}} \\
x_{\mathbb{S}_{i}}\\ 
y_{\mathbb{S}_{i}}\\ 
z_{\mathbb{S}_{i}}\\ 
\end{pmatrix}.
\label{aqn080}
\end{equation}
Thus, for sectors \(\mathbb{S}_{15}\) and \(\mathbb{S}_{16}\), it was determined that

\begin{equation}
\left|\beta_{\mathbb{S}_{15}}\right|=\begin{pmatrix}
8.049*10^{-16}\\
0.150 \\
7.246\\ 
4.916\\ 
2.304\\ 
\end{pmatrix}
\label{aqn090}
\end{equation}
and
\begin{equation}
\left|\beta_{\mathbb{S}_{16}}\right|=\begin{pmatrix}
6.661*10^{-16}\\
0.125 \\
9.459\\ 
6.665\\ 
3.343\\ 
\end{pmatrix}.
\label{aqn100}
\end{equation}
From \eqref{aqn090} and \eqref{aqn100}, it is possible to construct the money function \eqref{aqn040} and the partition function \eqref{eqn40}.

\begin{align}
    	Z\left(T, \bar{\lambda},\bar{\Lambda}\right)&=  \int_{\bar{\lambda}}\exp\left[-\frac{m_{\mathbb{S}_{i}}(\bar{\lambda},\bar{\Lambda})	}{T}\right]d\bar{\lambda}
\notag\\
	&=\int_{0}^{\infty}\exp\left[-\frac{k_{\mathbb{S}_{i}}}{T}\lambda^{v_{\mathbb{S}_{i}}}_{\mathbb{S}_{i}}\pi_{\mbox{\tiny{IPP}}} ^{x_{\mathbb{S}_{i}}}\pi_{\mbox{\tiny{IPC}}}^{y_{\mathbb{S}_{i}}}\lambda_{\mbox{\tiny{TRM}}}^{z_{\mathbb{S}_{i}}}\right]d\lambda_{\mathbb{S}_{i}}
&\hspace{0.3cm}
\notag\\
	&=\left[-\frac{k_{\mathbb{S}_{i}}}{T}\lambda^{v_{\mathbb{S}_{i}}}_{\mathbb{S}_{i}}\pi_{\mbox{\tiny{IPP}}} ^{x_{\mathbb{S}_{i}}}\pi_{\mbox{\tiny{IPC}}}^{y_{\mathbb{S}_{i}}}\lambda_{\mbox{\tiny{TRM}}}^{z_{\mathbb{S}_{i}}} \right]^{-1/v_{\mathbb{S}_{i}}}\Gamma\left[1+\frac{1}{v_{\mathbb{S}_{i}}}\right].
&\hspace{0.3cm}
\label{aqn110}
\end{align}
Once the partition function has been determined, the remaining thermodynamic properties can be derived. In particular, the free money function \eqref{eqn220} is given by
\begin{equation}
f_{\mathbb{S}_{i}}=-T\ln\left|\left[-\frac{k_{\mathbb{S}_{i}}}{T}\lambda^{v_{\mathbb{S}_{i}}}_{\mathbb{S}_{i}}\pi_{\mbox{\tiny{IPP}}} ^{x_{\mathbb{S}_{i}}}\pi_{\mbox{\tiny{IPC}}}^{y_{\mathbb{S}_{i}}}\sigma_{\mbox{\tiny{TRM}}}^{z_{\mathbb{S}_{i}}} \right]^{-1/v_{\mathbb{S}_{i}}}\Gamma\left[1+\frac{1}{v_{\mathbb{S}_{i}}} \right]\right|.
\label{aqn120}
\end{equation}
The entropy \( S_{\mathbb{S}_{i}} \) follows from Eq.~\eqref{eqn230}:
\begin{equation}
S_{\mathbb{S}_{i}}= \frac{1}{v_{\mathbb{S}_{i}}}+\ln\left|\left[-\frac{k_{\mathbb{S}_{i}}}{T}\lambda^{v_{\mathbb{S}_{i}}}_{\mathbb{S}_{i}}\pi_{\mbox{\tiny{IPP}}} ^{x_{\mathbb{S}_{i}}}\pi_{\mbox{\tiny{IPC}}}^{y_{\mathbb{S}_{i}}}\lambda_{\mbox{\tiny{TRM}}}^{z_{\mathbb{S}_{i}}} \right]^{-1/v_{\mathbb{S}_{i}}}\Gamma\left[1+\frac{1}{v_{\mathbb{S}_{i}}} \right]\right|.
\label{aqn130}
\end{equation}
The average money per agent in sector \(\mathbb{S}_{i}\) from Eq.~\eqref{eqn210}:

\begin{equation}
\left\langle m_{\mathbb{S}_{i}}\right\rangle=\frac{T}{v_{\mathbb{S}_{i}}},
\label{aqn140}
\end{equation}
and the heat capacity:

\begin{equation}
C_{\mathbb{S}_{i}}=\frac{1}{v_{\mathbb{S}_{i}}}
\label{aqn150}
\end{equation}
A first approximation to understanding the economic dynamics among sectors of the CSDB from a thermodynamic perspective is to assume that sectors \(\mathbb{S}_{i}\) and \(\mathbb{S}_{j}\) interact only with each other and not with the remaining sectors comprising the CSDB. This implies treating the subsystem as adiabatic. Consequently, each sector is characterized by its heat content \(Q_{\mathbb{S}_{i}}\) and its economic temperature \(T_{\mathbb{S}_{i}}\). CSDB-reported data indicate that the economic temperatures of different sectors are distinct, making it plausible to assume \(T_{\mathbb{S}_{15}} > T_{\mathbb{S}_{16}}\). Under these conditions, a heat-like transfer between the sectors—denoted \(Q_{\mathbb{S}_{15}} \Longleftrightarrow Q_{\mathbb{S}_{16}}\)—must occur. According to the second law of thermodynamics,
\begin{equation}
\Delta Q_{\mathbb{S}_{i}}=\int^{S_{f,\mathbb{S}_{i}}}_{{S_{o,\mathbb{S}_{i}}}}T_{\mathbb{S}_{i}}\left(S_{\mathbb{S}_{i}}\right)dS_{\mathbb{S}_{i}}.
\label{aqn160}
\end{equation}
From \eqref{aqn130} and \eqref{aqn160}, the thermal exchange \(\Delta Q_{\mathbb{S}_{i}}\) can be estimated:
\begin{equation}
\Delta Q_{\mathbb{S}_{i}}=\frac{k_{\mathbb{S}_{i}}\lambda^{v_{\mathbb{S}_{i}}}_{\mathbb{S}_{i}}\pi_{\mbox{\tiny{IPP}}}^{x_{\mathbb{S}_{i}}}\pi_{\mbox{\tiny{IPC}}}^{y_{\mathbb{S}_{i}}}\lambda^{z_{\mathbb{S}_{i}}}_{\mbox{\tiny{TRM}}}}{\Gamma\left[1+\frac{1}{v_{\mathbb{S}_{i}}} \right]^{v_{\mathbb{S}_{i}}}}\left(\frac{e^{v_{\mathbb{S}_{i}}-1}}{v_{\mathbb{S}_{i}}}\right)\Delta S_{\mathbb{S}_{i}}
\label{aqn170bb}
\end{equation}
Consequently, to estimate \eqref{aqn170bb}, we impose that the sectors form an isolated system, such that agents within each sector can only transact with agents of the other sector under consideration. Additionally, we assume that the total amount of money remains constant over a time interval \(\Delta t\). Therefore, the energy (money) transfer satisfies \(\Delta Q_{\mathbb{S}_{i}} = -\Delta Q_{\mathbb{S}_{j}}\).

\begin{figure}[ht]
\centering
		\includegraphics[width=0.45\textwidth]{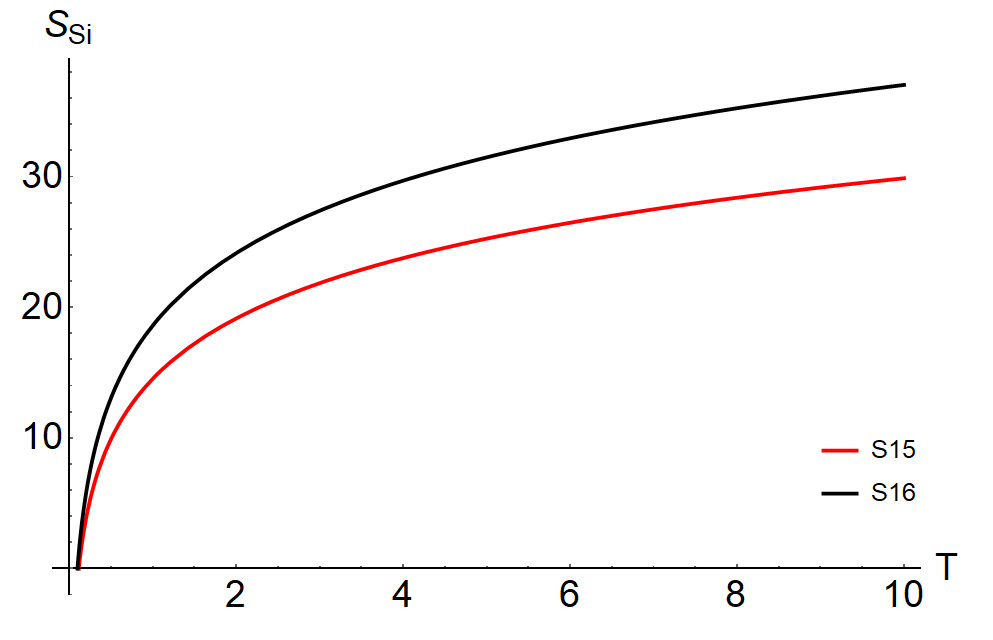}
\caption{Comparison of the entropies of sectors \(\mathbb{S}_{15}\) and \(\mathbb{S}_{16}\) in the CSDB.}
\label{QG0}
\end{figure}

Figure \ref{QG0} presents a comparison of the entropies of sectors \(\mathbb{S}_{15}\) and \(\mathbb{S}_{16}\) in the CSDB. It is observed that \(S_{\mathbb{S}_{15}} < S_{\mathbb{S}_{16}}\), implying that sector \(\mathbb{S}_{15}\)—corresponding to gambling and betting—is more organized and manages information more efficiently than sector \(\mathbb{S}_{16}\), which encompasses recreational and sports activities. Moreover, \(\mathbb{S}_{15}\) is subject to stricter governmental regulation.
 
\begin{figure}[ht]
\centering
		\includegraphics[width=0.45\textwidth]{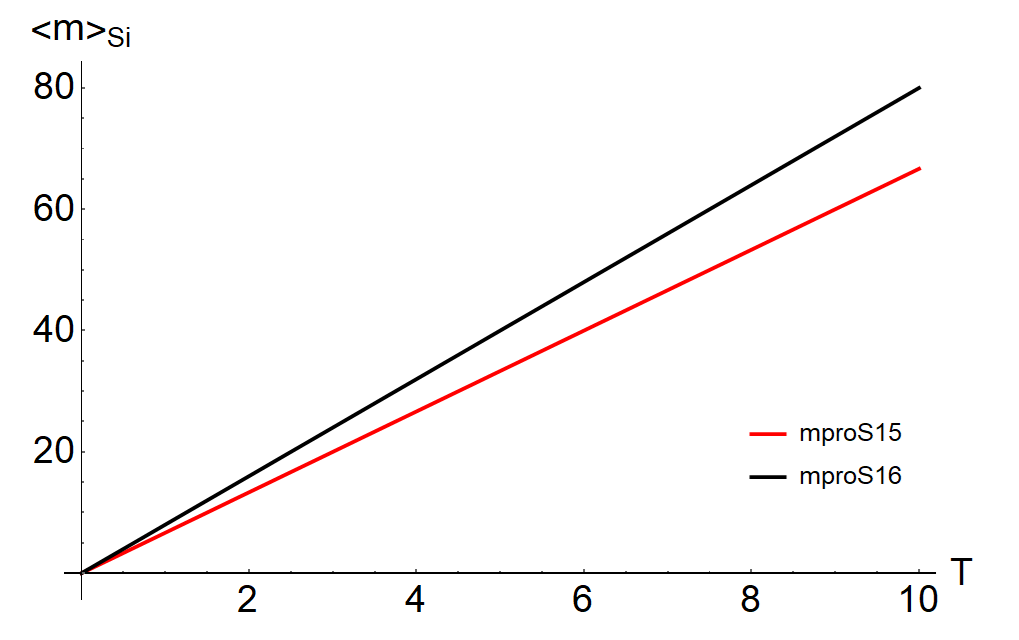}
\caption{Comparison of the average money per agent in sectors \(\mathbb{S}_{15}\) and \(\mathbb{S}_{16}\) of the CSDB.}
\label{QG10}
\end{figure}

Figure~\ref{QG10} shows a comparison of the average money per agent in sectors \(\mathbb{S}_{15}\) and \(\mathbb{S}_{16}\) of the CSDB. It is observed that \(\left\langle m_{\mathbb{S}_{16}} \right\rangle > \left\langle m_{\mathbb{S}_{15}} \right\rangle\), which can be interpreted as indicating that agents in sector \(\mathbb{S}_{16}\) require more monetary resources to carry out recreational and sports activities than agents in the gambling and betting market (\(\mathbb{S}_{15}\)). A reliable statistic on the number of market agents per sector could not be established, as this information is confidential.

\begin{figure}[ht]
\centering
		\includegraphics[width=0.45\textwidth]{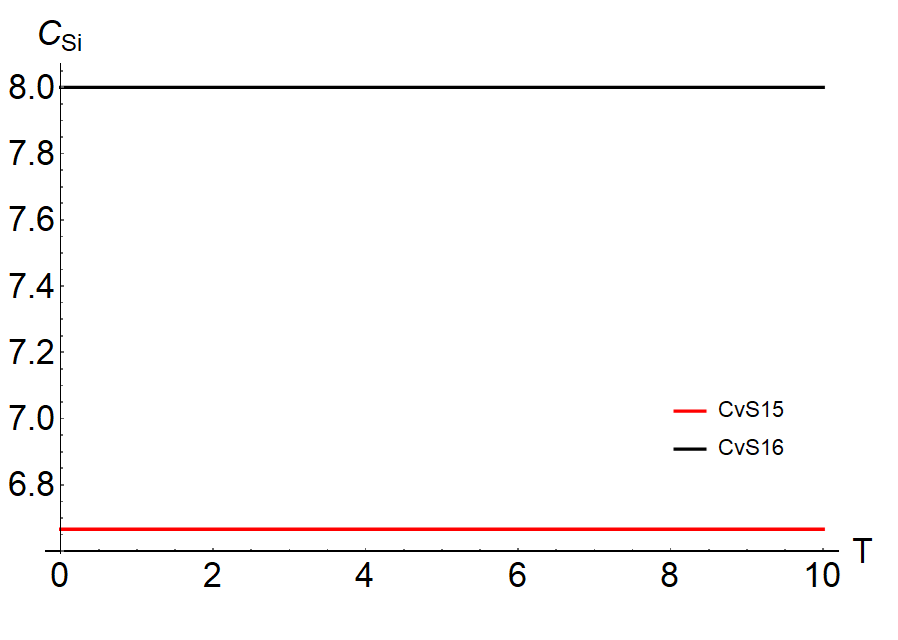}
\caption{Comparison of the heat capacities of sectors \(\mathbb{S}_{15}\) and \(\mathbb{S}_{16}\) in the CSDB.}
\label{QG20}
\end{figure}

Figure~\ref{QG20} presents a comparison of the heat capacities of sectors \(\mathbb{S}_{15}\) and \(\mathbb{S}_{16}\) in the CSDB. According to the thermodynamic definition, heat capacity measures the amount of energy required to raise a system’s temperature by one kelvin \cite{zemansky1968heat}. The results show that \(C_{\mathbb{S}_{16}} > C_{\mathbb{S}_{15}}\), implying that sector \(\mathbb{S}_{16}\) (recreational and sports activities) requires more activation energy—i.e., greater investment—to sustain its business dynamics. This includes substantial expenditures on specialized sports infrastructure such as stadiums, training centers, and specialized equipment.  

In contrast, the lower heat capacity of sector \(\mathbb{S}_{15}\) (gambling and betting) indicates that it demands significantly less investment in physical infrastructure or technology. This is further facilitated by the widespread availability of communication devices (e.g., smartphones, tablets, and other internet-connected gadgets), declining costs of 4G/5G connectivity, and expanded network coverage, which collectively ensure broad and low-cost access to the betting market across the population.

\begin{figure}[ht]
\centering
		\includegraphics[width=0.45\textwidth]{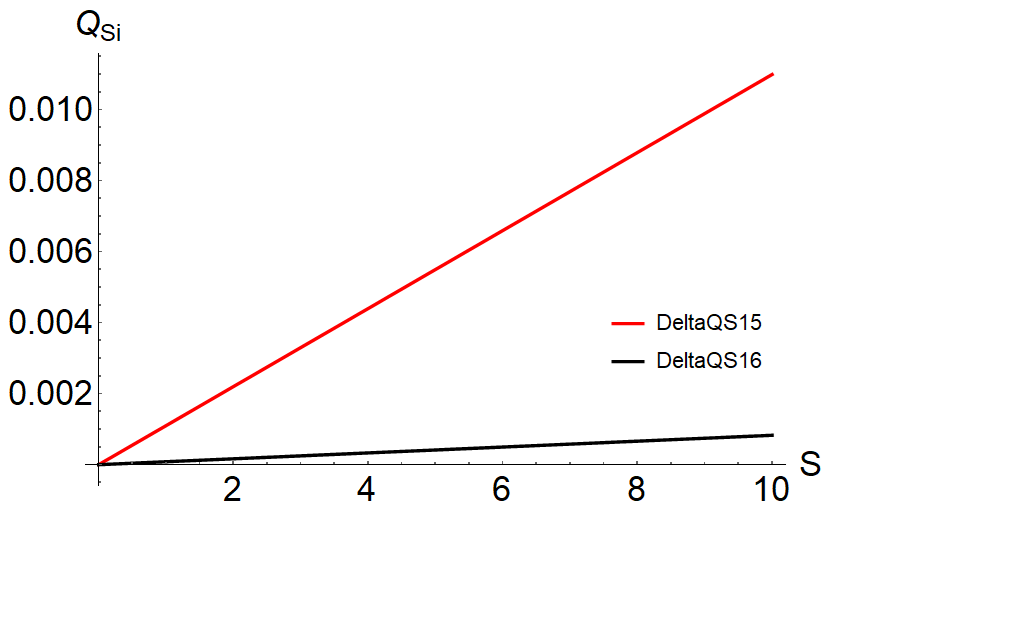}
\caption{Comparison of \(\Delta Q\) for sectors \(\mathbb{S}_{15}\) and \(\mathbb{S}_{16}\) in the CSDB..}
\label{QG30}
\end{figure}

Figure \ref{QG30} presents a comparison of \(\Delta Q\) for sectors \(\mathbb{S}_{15}\) and \(\mathbb{S}_{16}\) in the CSDB. It is evident that \(\mathbb{S}_{15}\) thermalizes more rapidly than \(\mathbb{S}_{16}\), with \(Q_{\mathbb{S}_{15}} > Q_{\mathbb{S}_{16}}\). This indicates that the gambling and betting sector exhibits more dynamic economic activity compared to the recreational and sports sector. Consequently, a heat-like transfer of economic energy must occur from \(\mathbb{S}_{15}\) to \(\mathbb{S}_{16}\), i.e., \(Q_{\mathbb{S}_{15}} \longrightarrow Q_{\mathbb{S}_{16}}\), in accordance with thermodynamic principles governing energy flow from higher- to lower-temperature systems.

\section{Econophysics Approach to Sectoral Dynamics}
Consider a more realistic approximation: given that the CSDB comprises 17 complementary economic activities, and that 5 of these account for the largest share of output, we describe the money function \eqref{eqn40} as \cite{Quevedo2011}:
\begin{equation}
 m_{\mathbb{S}_{i}}(\bar{\lambda},\bar{\Lambda})=  k_{\mathbb{S}_{i}} \left[\lambda^{v_{\mathbb{S}_{i}}}_{\mathbb{S}_{i}}+ \ln \left|\pi_{\mbox{\tiny{IPP}}} ^{x_{\mathbb{S}_{i}}}\pi_{\mbox{\tiny{IPC}}}^{y_{\mathbb{S}_{i}}}\lambda_{\mbox{\tiny{TRM}}}^{z_{\mathbb{S}_{i}}}\right|\right].
\label{aqn170}
\end{equation}
Then, the partition function becomes:
\begin{widetext} 
\begin{align}
    	Z\left(T, \bar{\lambda},\bar{\Lambda}\right)&=  \int_{\bar{\lambda}} \int_{\bar{\Lambda}}\exp\left[\frac{m_{\mathbb{S}_{i}}(\bar{\lambda},\bar{\Lambda})}{T}\right]d\bar{\lambda}d\bar{\Lambda}
\notag\\
	&=\int^{\infty}_{0}\exp\left[-\frac{k_{\mathbb{S}_{i}}\lambda^{v_{\mathbb{S}_{i}}}_{\mathbb{S}_{i}}}{T}\right]d\lambda_{\mathbb{S}_{i}} \int^{X}_{0}\int^{Y}_{0}\int^{Z}_{1} \pi_{\mbox{\tiny{IPP}}} ^{\frac{-k_{\mathbb{S}_{i}}  x_{\mathbb{S}_{i}}}{T}} \pi_{\mbox{\tiny{IPC}}} ^{\frac{-k_{\mathbb{S}_{i}}  y_{\mathbb{S}_{i}}}{T}}  \lambda_{\mbox{\tiny{TRM}}} ^{\frac{-k_{\mathbb{S}_{i}}  z_{\mathbb{S}_{i}}}{T}} d \pi_{\mbox{\tiny{IPP}}} d \pi_{\mbox{\tiny{IPC}}} d \lambda_{\mbox{\tiny{TRM}}}
&\hspace{0.3cm}
\notag\\
	&=-\frac{T^{3}\Delta X\Delta Y \Delta Z}{D_{\mathbb{S}_{i}}}\left(\frac{k_{\mathbb{S}_{i}}}{T}\right)^{-1/v_{\mathbb{S}_{i}}}\Gamma_{v_{\mathbb{S}_{i}}},
&\hspace{0.3cm}
\label{aqn180}
\end{align}
\end{widetext}
where, for \eqref{aqn180}, the following auxiliary variables have been defined:
\begin{equation}
A_{\mathbb{S}_{i}}=\frac{k_{\mathbb{S}_{i}}  x_{\mathbb{S}_{i}}}{T}\,\,B_{\mathbb{S}_{i}}=\frac{k_{\mathbb{S}_{i}}  y_{\mathbb{S}_{i}}}{T},\,\,C_{\mathbb{S}_{i}}=\frac{k_{\mathbb{S}_{i}}  z_{\mathbb{S}_{i}}}{T},
\label{aqn190}
\end{equation}
\begin{equation}
\Delta X=X^{1-A_{\mathbb{S}_{i}}}-X_{0}^{1-A_{\mathbb{S}_{i}}},\,\,\Delta Y=Y^{1-B_{\mathbb{S}_{i}}}-Y_{0}^{1-B_{\mathbb{S}_{i}}},\,\,
\label{aqn200}
\end{equation}
	\[\Delta Z=1-Z_{0}^{1-C_{\mathbb{S}_{i}}}
\]
\begin{equation}
\Gamma_{v_{\mathbb{S}_{i}}}=\Gamma\left[1+\frac{1}{v_{\mathbb{S}_{i}}}\right]
\label{aqn210}
\end{equation}
\begin{equation}
D_{\mathbb{S}_{i}}=\left(T-k_{\mathbb{S}_{i}}  x_{\mathbb{S}_{i}}\right)\left(T-k_{\mathbb{S}_{i}}  y_{\mathbb{S}_{i}}\right)\left(T-k_{\mathbb{S}_{i}}  z_{\mathbb{S}_{i}}\right).
\label{aqn220}
\end{equation}
The free money function is obtained from \eqref{eqn210}, namely:
\begin{equation}
\left\langle m\right\rangle_{\mathbb{S}_{i}}=T^{2}\frac{\partial}{\partial T}\ln\left|-\frac{T^{3}\Delta X\Delta Y \Delta Z}{D_{\mathbb{S}_{i}}}\left(\frac{k_{\mathbb{S}_{i}}}{T}\right)^{-1/v_{\mathbb{S}_{i}}}\Gamma_{v_{\mathbb{S}_{i}}} \right|.
\label{aqn230}
\end{equation}
The heat capacity is obtained as in \eqref{eqn241}:
\begin{equation}
C_{\mathbb{S}_{i}}=T\frac{\partial^{2}}{\partial T^{2}}\ln\left|-\frac{T^{3}\Delta X\Delta Y \Delta Z}{D_{\mathbb{S}_{i}}}\left(\frac{k_{\mathbb{S}_{i}}}{T}\right)^{-1/v_{\mathbb{S}_{i}}}\Gamma_{v_{\mathbb{S}_{i}}} \right|
\label{aqn240bb}
\end{equation}
\begin{figure}[ht]
\centering
		\includegraphics[width=0.45\textwidth]{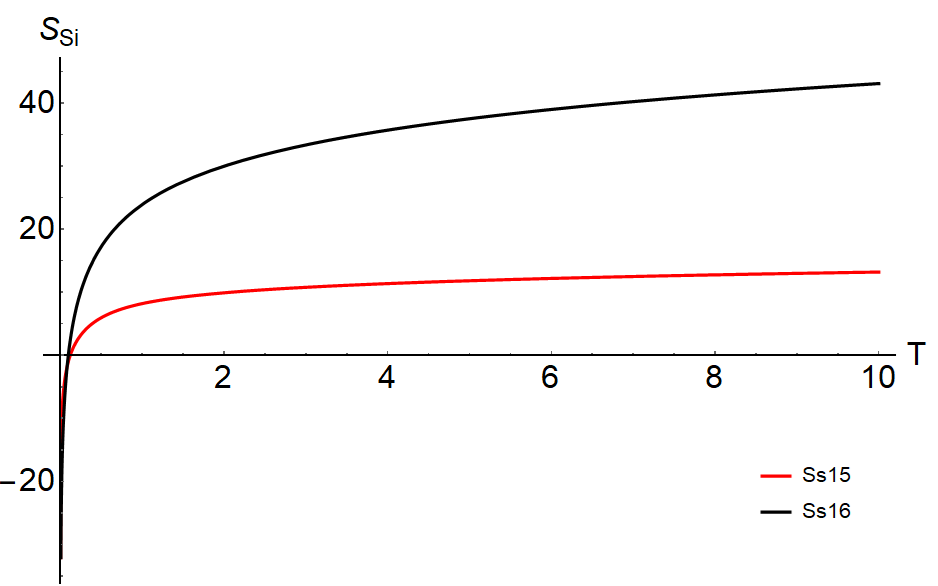}
\caption{Comparison of the entropies of sectors \(\mathbb{S}_{15}\) and \(\mathbb{S}_{16}\) in the CSDB under the extended approximation.}
\label{QG40}
\end{figure}
Figure~\ref{QG40} presents a comparison of the entropies of sectors \(\mathbb{S}_{15}\) and \(\mathbb{S}_{16}\) in the CSDB under the extended approximation. It is observed that \(S_{\mathbb{S}_{15}} < S_{\mathbb{S}_{16}}\), a behavior consistent with that reported in Figure~\ref{QG0}. This indicates a higher degree of disorder in sector \(\mathbb{S}_{16}\), which is attributable to its intrinsic nature—recreational and sports activities—characterized by greater heterogeneity and less centralized coordination.  

In contrast, the lower entropy of sector \(\mathbb{S}_{15}\) reflects a higher level of self-organization and more efficient information management. Contributing factors include the extensive use of digital technologies for bet processing, real-time transaction monitoring, and statistical analytics. Furthermore, this sector is subject to strict regulation and oversight by the Colombian government through Coljuegos \cite{coljuegos2023}, which enforces operational standards and enhances systemic coherence.

\begin{figure}[ht]
\centering
		\includegraphics[width=0.45\textwidth]{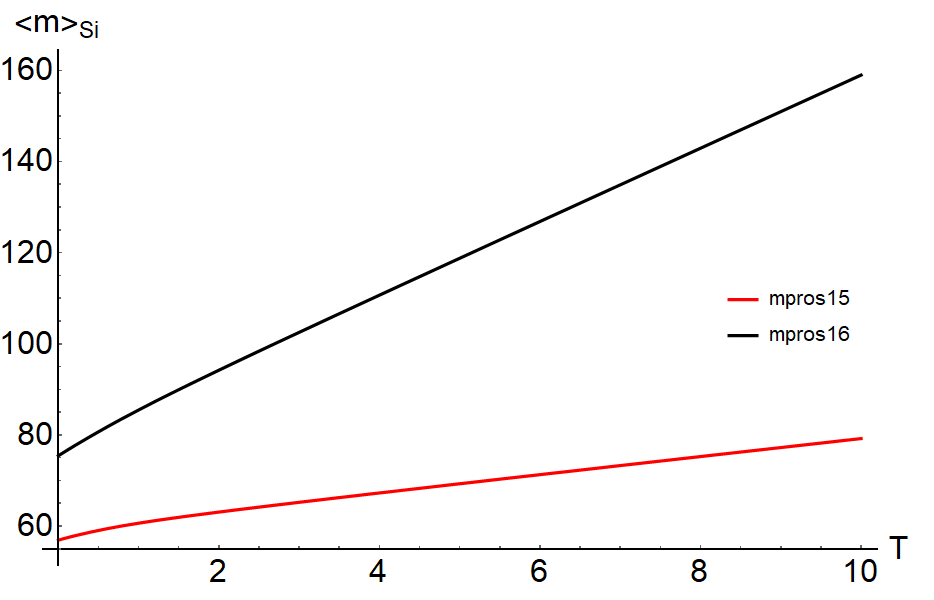}
\caption{Comparison of the average money per agent in sectors \(\mathbb{S}_{15}\) and \(\mathbb{S}_{16}\) of the CSDB.}
\label{QG50}
\end{figure}


Figure~\ref{QG50} shows a comparison of the average money per agent in sectors \(\mathbb{S}_{15}\) and \(\mathbb{S}_{16}\) of the CSDB. It is observed that \(\left\langle m_{\mathbb{S}_{16}} \right\rangle > \left\langle m_{\mathbb{S}_{15}} \right\rangle\), a behavior consistent with that reported in Figure~\ref{QG10}. This reinforces the earlier argument: significantly higher investments are required to operate in sector \(\mathbb{S}_{16}\), primarily due to the need for specialized recreational and sports infrastructure, specialized equipment, and consumables—many of which are not produced domestically and must be imported.

In contrast, sector \(\mathbb{S}_{15}\) demands far less physical infrastructure, equipment, or consumables. Access to online betting platforms is largely facilitated through widely available mobile devices (smartphones, tablets) and internet connectivity, drastically lowering entry and operational barriers \cite{dane2024}.
\begin{figure}[ht]
\centering
		\includegraphics[width=0.45\textwidth]{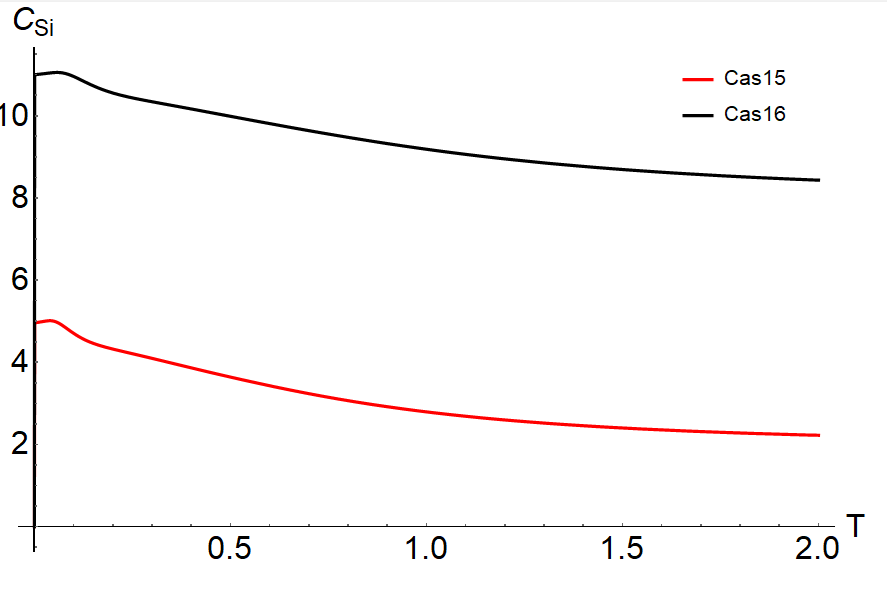}
\caption{Comparison of the heat capacities of sectors \(\mathbb{S}_{15}\) and \(\mathbb{S}_{16}\) in the CSDB.}
\label{QG60}
\end{figure}
Figure \ref{QG60} presents a comparison of the heat capacities of sectors \(\mathbb{S}_{15}\) and \(\mathbb{S}_{16}\) in the CSDB, showing behavior consistent with that reported in Figure \ref{QG20}. It is observed that \(C_{\mathbb{S}_{16}} > C_{\mathbb{S}_{15}}\). In both cases, the heat capacities exhibit a maximum and subsequently decrease with increasing economic temperature \(T\). Although an explicit expansion of \eqref{aqn240bb} is lengthy and cumbersome for the scope of this study, it is worth noting that \(C_{\mathbb{S}_{i}}\) displays singular points at specific temperatures, which can be interpreted as phase transitions in the economic system \cite{Quevedo2011,Quevedo2023,Quevedo2007}.

Furthermore, when a temperature gradient exists between the two sectors, it is possible to estimate the heat-like transfer \(Q\) between \(\mathbb{S}_{15}\) and \(\mathbb{S}_{16}\) using their respective heat capacities, following the thermodynamic analogy established in classical heat transfer theory \cite{zemansky1968heat}.
\begin{equation}
C_{\mathbb{S}_{i}}=\frac{dQ_{\mathbb{S}_{i}}}{dT_{\mathbb{S}_{i}}}.
\label{aqn240}
\end{equation}
Therefore
\begin{widetext} 
\begin{align}
\Delta Q_{\mathbb{S}_{i}}&=\int^{T_{f,\mathbb{S}_{i}}}_{{T_{o,\mathbb{S}_{i}}}}C_{\mathbb{S}_{i}} dT_{\mathbb{S}_{i}} 
\notag\\
	&=\int^{T_{f,\mathbb{S}_{i}}}_{{T_{o,\mathbb{S}_{i}}}}T\frac{\partial^{2}}{\partial T^{2}}\ln\left|-\frac{T^{3}\Delta X\Delta Y \Delta Z}{D_{\mathbb{S}_{i}}}\left(\frac{k_{\mathbb{S}_{i}}}{T}\right)^{-1/v_{\mathbb{S}_{i}}}\Gamma_{v_{\mathbb{S}_{i}}} \right|dT_{\mathbb{S}_{i}},
\label{aqn250}
\end{align}
\end{widetext}
where, for \eqref{aqn250}, we consider \( S_{\text{Total}} = S_{\mathbb{S}_{15}} + S_{\mathbb{S}_{16}} \) and impose \( \Delta Q_{\mathbb{S}_{i}} = -\Delta Q_{\mathbb{S}_{j}} \). This implies that the two economic sectors behave adiabatically with respect to all other sectors comprising the CSDB. Additionally, the total amount of money \( M \) is conserved, and the number of agents in each sector remains fixed \cite{Quevedo2011, Chakrabarti2013, dragulescu2000statistical, DragulescuYakovenko2003}.

\begin{figure}[ht]
\centering
 		\includegraphics[width=0.45\textwidth]{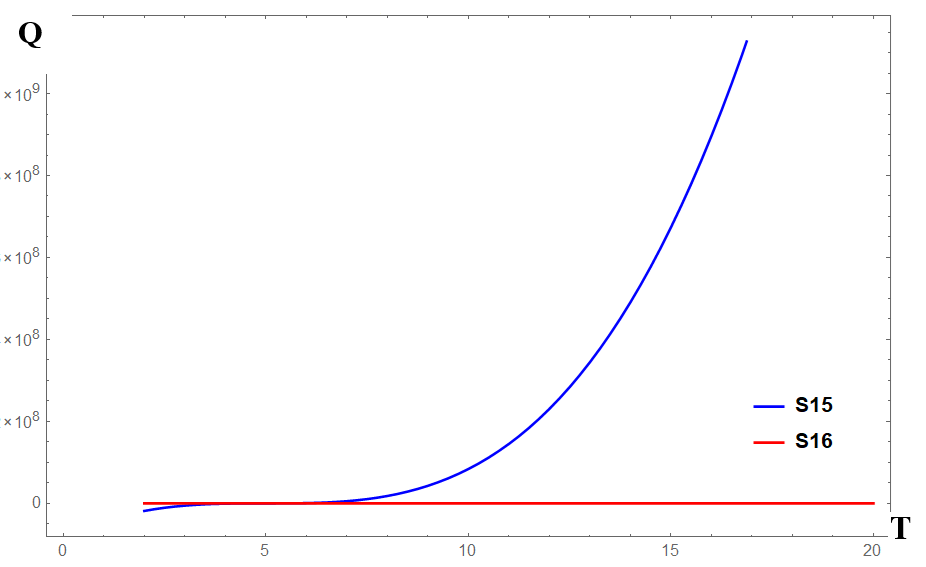}
\caption{Comparison of \(\Delta Q\) for sectors \(\mathbb{S}_{15}\) and \(\mathbb{S}_{16}\) in the CSDB.}
\label{QG70}
\end{figure}
Figure~\ref{QG70} presents \(\Delta Q\) for sectors \(\mathbb{S}_{15}\) and \(\mathbb{S}_{16}\) in the CSDB. It is observed that \(\Delta Q_{\mathbb{S}_{15}} > \Delta Q_{\mathbb{S}_{16}}\). This is interpreted as follows: the two sectors form an adiabatic subsystem isolated from the rest of the CSDB, and a temperature gradient exists between them. Consequently, a heat-like flow occurs from \(\mathbb{S}_{15}\) to \(\mathbb{S}_{16}\), i.e., \(\Delta Q_{\mathbb{S}_{15}} \longrightarrow \Delta Q_{\mathbb{S}_{16}}\), implying that sector \(\mathbb{S}_{15}\) effectively supplies economic resources to sector \(\mathbb{S}_{16}\).

The fact that \(\Delta Q > 0\) indicates that this inter-sectoral transfer of heat (i.e., money) is analogous to a transport process in thermodynamics, where energy flows spontaneously from a higher-temperature to a lower-temperature system \cite{landau1988fisica}.

\section{Geometrothermodynamic Approximation to the CSDB}
Let the money function \eqref{eqn40} be extended to include the number of firms in sector \(\mathbb{S}_{i}\):
\begin{equation}
 m_{\mathbb{S}_{i}}(\bar{\lambda},\bar{\Lambda})=  k_{\mathbb{S}_{i}} \left[\lambda^{v_{\mathbb{S}_{i}}}_{\mathbb{S}_{i}}+ \ln \left|\pi_{\mbox{\tiny{IPP}}} ^{x_{\mathbb{S}_{i}}}\pi_{\mbox{\tiny{IPC}}}^{y_{\mathbb{S}_{i}}}\lambda_{\mbox{\tiny{TRM}}}^{z_{\mathbb{S}_{i}}}\right|+\ln \left|\frac{N_{\mathbb{S}_{i},f}}{N_{\mathbb{S}_{i},o}}\right|\right],
\label{aqn260}
\end{equation}
where \( N_{\mathbb{S}_{i},0} \) is the number of firms in sector \( \mathbb{S}_{i} \) at the initial time \( t_{0} = 2018 \), and \( N_{\mathbb{S}_{i},f} \) is the number of firms at time \( t \). The partition function then becomes:

\begin{equation}
Z\left(T, \bar{\lambda},\bar{\Lambda}\right)=\frac{T^{4 } \Delta N\Delta X\Delta Y \Delta Z}{D_{\mathbb{S}_{i}}}\left(\frac{k_{\mathbb{S}_{i}}}{T}\right)^{-1/v_{\mathbb{S}_{i}}}\Gamma_{v_{\mathbb{S}_{i}}},
\label{aqn270}
\end{equation}

where for \eqref{aqn270}, the following auxiliary variables have been defined:
\begin{equation}
A_{\mathbb{S}_{i}}=\frac{k_{\mathbb{S}_{i}}  x_{\mathbb{S}_{i}}}{T}\,\,B_{\mathbb{S}_{i}}=\frac{k_{\mathbb{S}_{i}}y_{\mathbb{S}_{i}}}{T},\,\,,
\label{aqn280}
\end{equation}

\begin{equation}
C_{\mathbb{S}_{i}}=\frac{k_{\mathbb{S}_{i}}  z_{\mathbb{S}_{i}}}{T},\,\,\Delta N=1-N_{\mathbb{S}_{i},o}^{1- \frac{k_{\mathbb{S}_{i}}  x_{\mathbb{S}_{i}}}{T}}
\label{aqn280a}
\end{equation}

\begin{equation}
D_{\mathbb{S}_{i}}=\left(T-k_{\mathbb{S}_{i}}\right)\left(T-k_{\mathbb{S}_{i}}  x_{\mathbb{S}_{i}}\right) \left(T-k_{\mathbb{S}_{i}}  y_{\mathbb{S}_{i}}\right)\left(T-k_{\mathbb{S}_{i}}  z_{\mathbb{S}_{i}}\right).
\label{aqn290}
\end{equation}
Then, the entropy is:
\begin{widetext} 
\begin{equation}
S_{\mathbb{S}_{i}}\left(T,N_{\mathbb{S}_{i}}\right)=\ln \left|\frac{T^{4 } \Delta N\Delta X\Delta Y \Delta Z}{D_{\mathbb{S}_{i}}}\left(\frac{k_{\mathbb{S}_{i}}}{T}\right)^{-1/v_{\mathbb{S}_{i}}}\Gamma_{v_{\mathbb{S}_{i}}}\right| +T \frac{\partial}{\partial T} \ln \left|\frac{T^{4 } \Delta N\Delta X\Delta Y \Delta Z}{D_{\mathbb{S}_{i}}}\left(\frac{k_{\mathbb{S}_{i}}}{T}\right)^{-1/v_{\mathbb{S}_{i}}}\Gamma_{v_{\mathbb{S}_{i}}}\right|,
\label{aqn300}
\end{equation}
\end{widetext} 

An explicit calculation of \eqref{aqn300} is too cumbersome for the scope of this study. Therefore, within the Geometrothermodynamics framework—where the thermodynamic potential depends only on the extensive variables of the system \cite{pineda2019geometrotermodinamica}—we take \(\Phi = S(T, N_{\mathbb{S}_{i}})\) with extensive coordinates \(E^{a} = \{T, N_{\mathbb{S}_{i}}\} = \{E^{1}, E^{2}\}\). This allows us to compute the metric tensor \eqref{gtd20} on the equilibrium manifold \(\mathcal{E}\).  

\begin{figure}[ht]
\centering
		\includegraphics[width=0.45\textwidth]{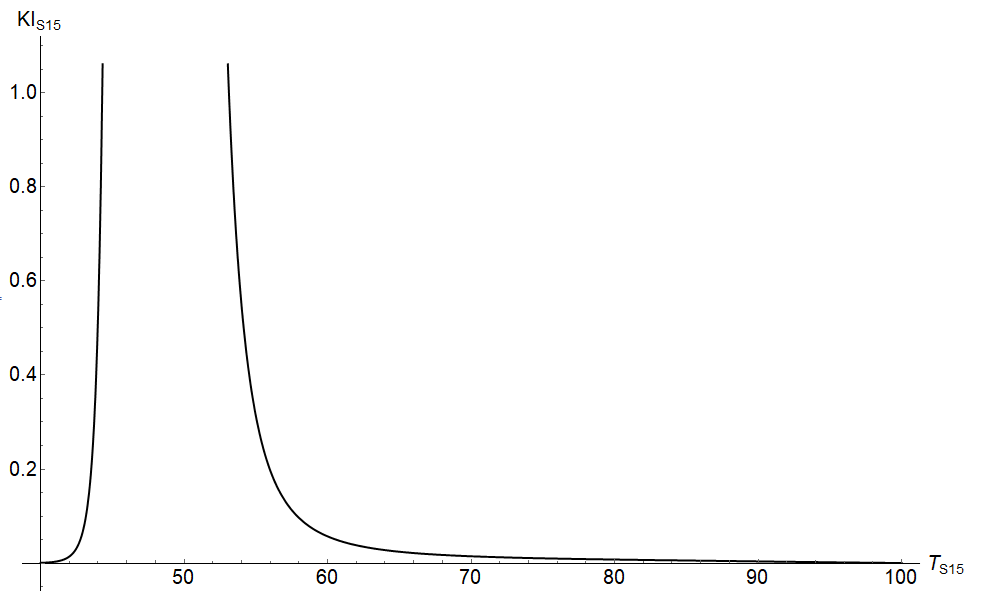}
\caption{Kretschmann scalar \( K^{I}_{\mathbb{S}_{15}} \) for sector \( \mathbb{S}_{15} \) with \( N_{\mathbb{S}_{15}} = \text{const} \).}
\label{QG80}
\end{figure}

\begin{figure}[ht]
\centering
		\includegraphics[width=0.45\textwidth]{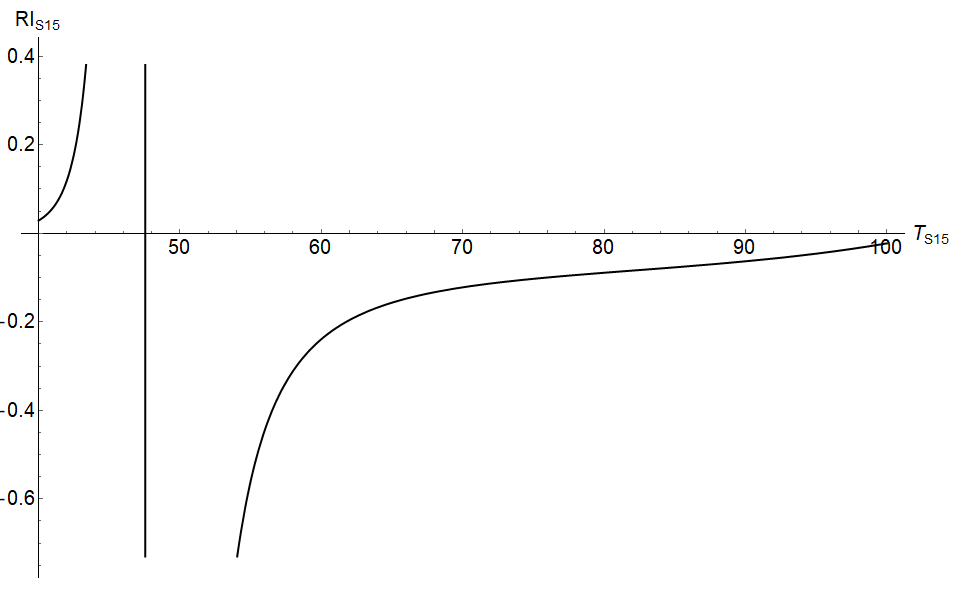}
\caption{Ricci scalar \( R^{I}_{\mathbb{S}_{15}} \) for sector \( \mathbb{S}_{15} \) with \( N_{\mathbb{S}_{15}} = \text{const} \).}
\label{QG90}
\end{figure}

A first-order Taylor series expansion of Eq.~\eqref{aqn300} in two variables yields:
\begin{widetext} 
\begin{equation}
S_{\mathbb{S}_{i}}\left(T,N_{\mathbb{S}_{i}}\right)\approx S\left(T_{o},N_{\mathbb{S}_{i},0} \right) + \left(T-T_{o}\right)\frac{\partial S\left(T_{o},N_{\mathbb{S}_{i},0}\right)}{\partial T}+ \left(N_{\mathbb{S}_{i}}-N_{\mathbb{S}_{i},o}\right) \frac{\partial S\left(T_{o},N_{\mathbb{S}_{i},0}\right)}{\partial N_{\mathbb{S}_{i}}}.
\label{aqn310}
\end{equation}
 \end{widetext} 
This enabled the computation of the metric \( g^{I} \) from Eq.~\eqref{gtd20}. Plots of the scalar curvature \( K^{I}_{\mathbb{S}_{15}} \) and Ricci scalar \( R^{I}_{\mathbb{S}_{15}} \) for \( N_{\mathbb{S}_{i}} = 8500 \) are shown in Figures \ref{QG80} and \ref{QG90}, respectively. Both curvature scalars diverge at \( T = 50 \), which is interpreted as a signature of a phase transition in the equilibrium space. It should be noted that the study period (2018–2023) includes the COVID-19 pandemic, a globally disruptive event that significantly impacted economic activity and likely triggered the observed critical behavior.

\begin{figure}[ht]
\centering
		\includegraphics[width=0.45\textwidth]{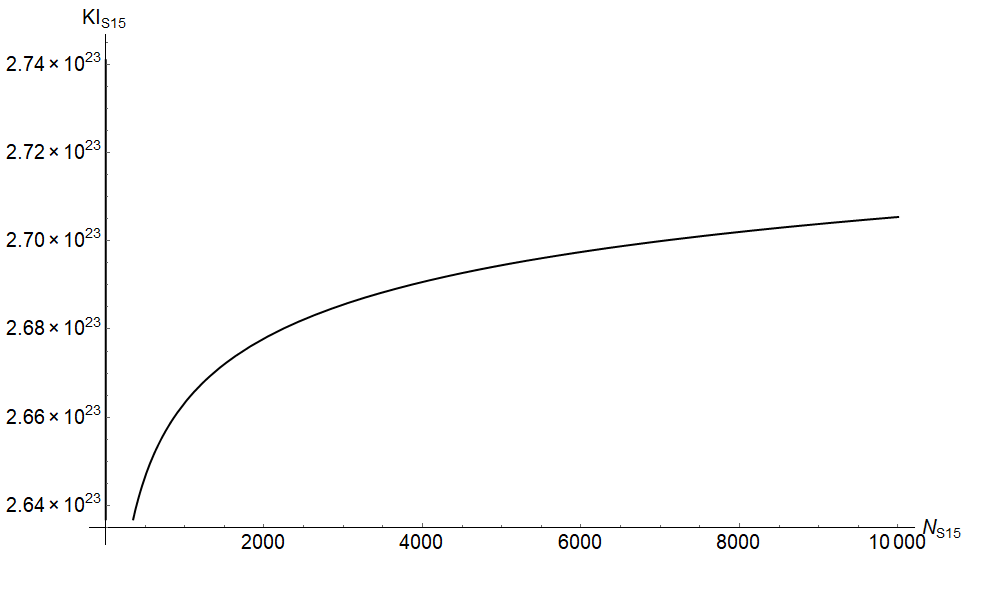}
\caption{Kretschmann scalar \( K^{I}_{\mathbb{S}_{15}} \) for sector \( \mathbb{S}_{15} \) with \( T = \text{const} \).}
\label{QG100}
\end{figure}

\begin{figure}[ht]
\centering
		\includegraphics[width=0.45\textwidth]{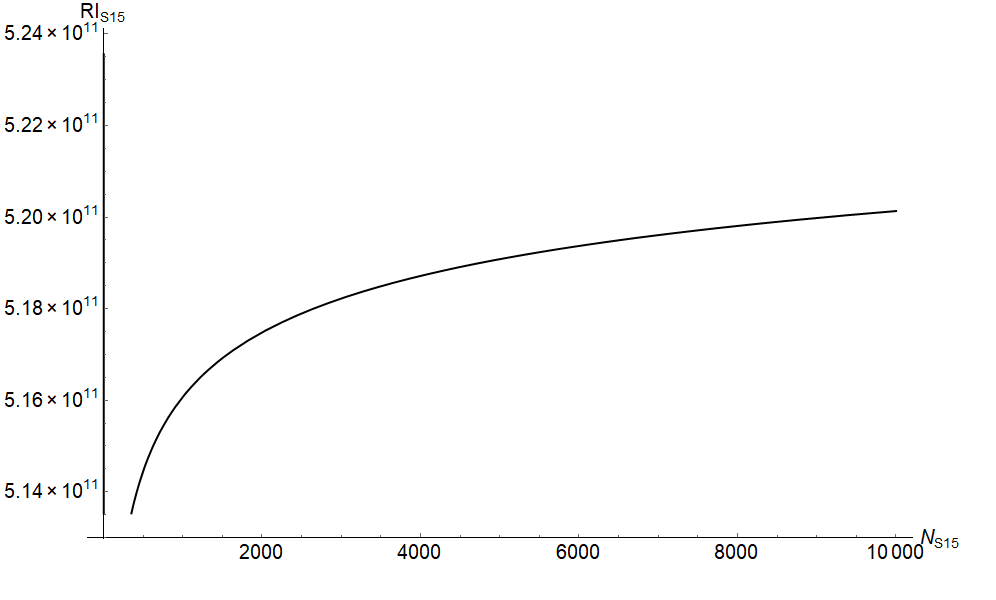}
\caption{Ricci scalar \( R^{I}_{\mathbb{S}_{15}} \) for sector \( \mathbb{S}_{15} \) with \( T = \text{const} \).}
\label{QG110}
\end{figure}

Plots of \( K^{I}_{\mathbb{S}_{15}} \) and \( R^{I}_{\mathbb{S}_{15}} \) at constant temperature (\( T = \text{const} \)) are shown in Figures~\ref{QG100} and~\ref{QG110}, respectively. The curvature scalars exhibit a singularity near the origin, which can be attributed to divergences in the heat capacity \eqref{aqn240bb} of the sector. For large values of \( N_{\mathbb{S}_{15}} \), both scalars increase monotonically after the pandemic period, reflecting a stabilization and growth phase in the sector’s thermodynamic structure.

This implies that the curvature scalars discussed above exhibit divergences—i.e., singular points on the equilibrium manifold \(\mathcal{E}\)—which should be interpreted as signatures of phase transitions, corresponding in this context to economic crises affecting sector \(\mathbb{S}_{15}\).

\begin{figure}[ht]
\centering
		\includegraphics[width=0.45\textwidth]{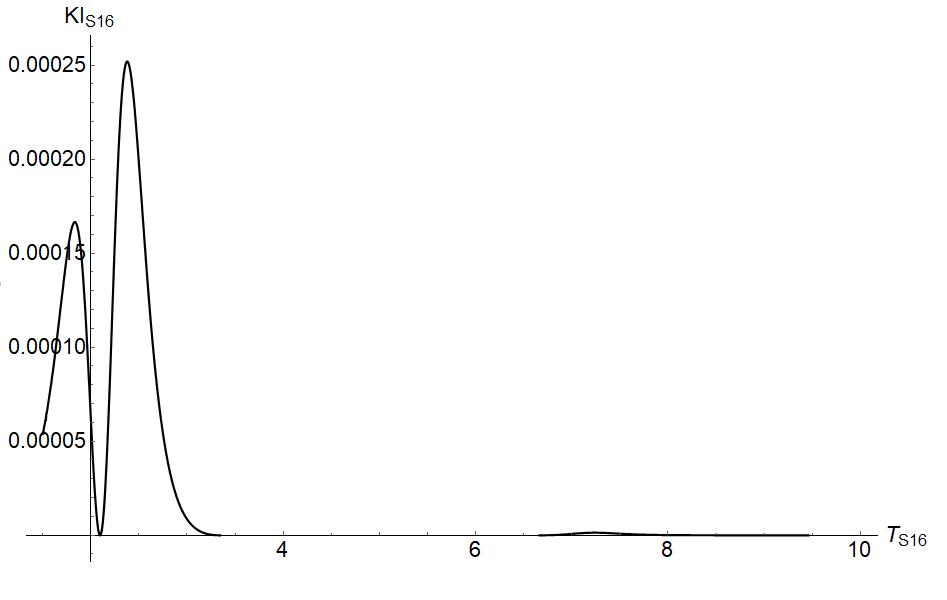}
\caption{Kretschmann scalar \( K^{I}_{\mathbb{S}_{16}} \) for sector \( \mathbb{S}_{16} \) with \( N_{\mathbb{S}_{16}} = \text{const} \).}
\label{QG120}
\end{figure}

\begin{figure}[ht]
\centering
		\includegraphics[width=0.45\textwidth]{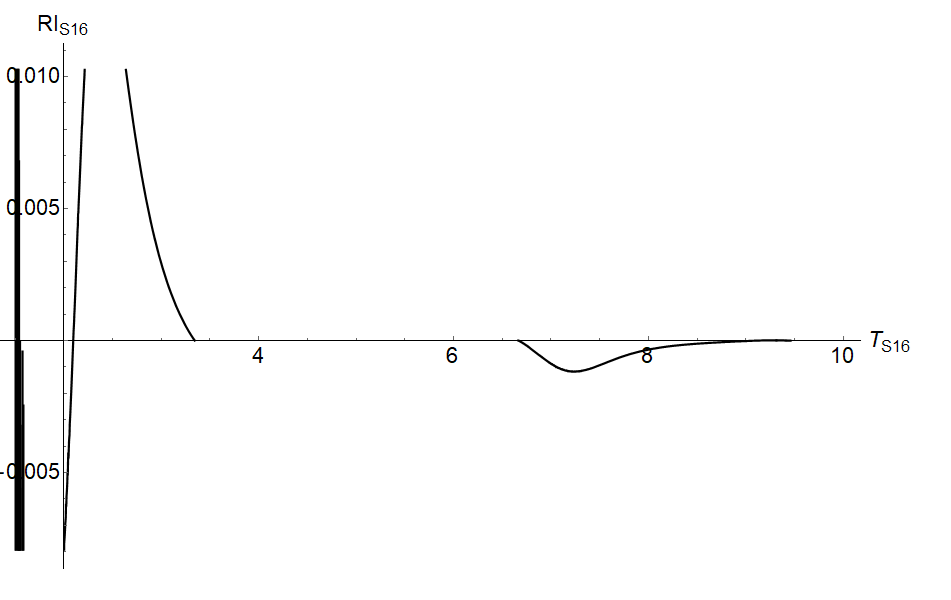}
\caption{Ricci scalar \( R^{I}_{\mathbb{S}_{16}} \) for sector \( \mathbb{S}_{16} \) with \( N_{\mathbb{S}_{16}} = \text{const} \).}
\label{QG130}
\end{figure}

Additionally, plots of \(K^{I}_{\mathbb{S}_{16}}\) and \(R^{I}_{\mathbb{S}_{16}}\) for a fixed number of firms \(N_{\mathbb{S}_{16}} = 9500\) are shown in Figures~\ref{QG120} and~\ref{QG130}, respectively. At low temperatures, the curvature scalars display fluctuations and subsequently flatten out, suggesting that the thermodynamic interactions within this sector are relatively weak. This behavior is consistent with the fact that the CSDB accounts for only \(1.1\%\) of Bogotá’s GDP, reflecting the sector’s limited macroeconomic weight and lower systemic coupling.

\begin{figure}[ht]
\centering
		\includegraphics[width=0.45\textwidth]{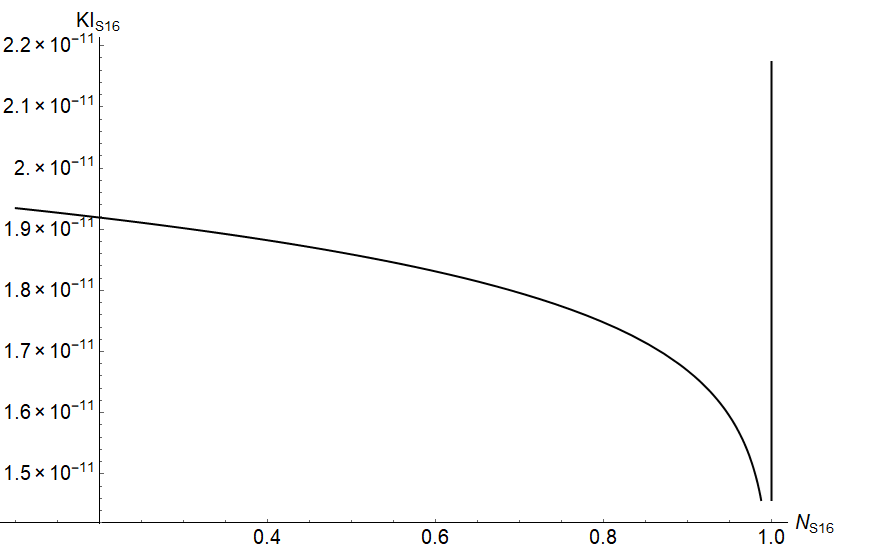}
\caption{Kretschmann scalar \( K^{I}_{\mathbb{S}_{16}} \) for sector \( \mathbb{S}_{16} \) with \( T = \text{const} \).}
\label{QG140}
\end{figure}

\begin{figure}[ht]
\centering
		\includegraphics[width=0.45\textwidth]{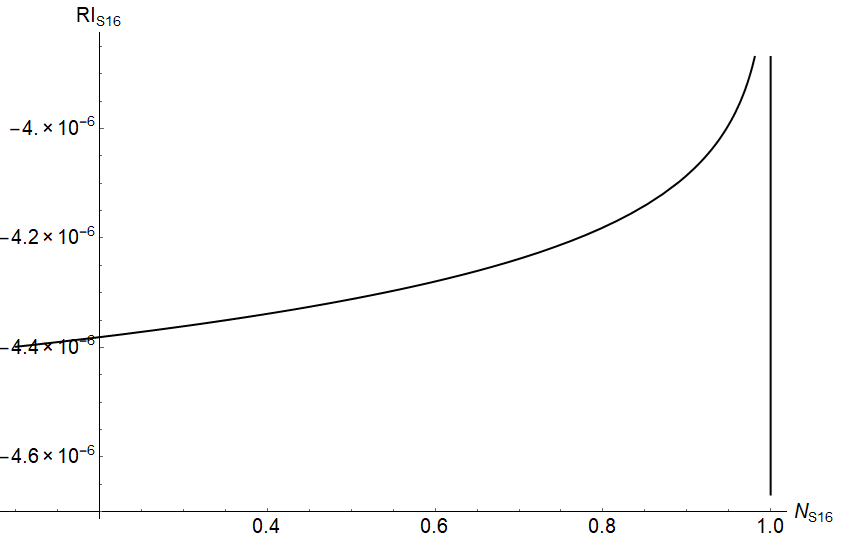}
\caption{Ricci scalar \( R^{I}_{\mathbb{S}_{16}} \) for sector \( \mathbb{S}_{16} \) with \( T = \text{const} \).}
\label{QG150}
\end{figure}

Plots of \( K^{I}_{\mathbb{S}_{16}} \) and \( R^{I}_{\mathbb{S}_{16}} \) at constant temperature (\( T = \text{const} \)) are shown in Figures~\ref{QG140} and~\ref{QG150}, respectively. It is observed that for \( N_{\mathbb{S}_{16}} = 1 \), both curvature scalars diverge, likely indicating singularities on the equilibrium manifold \(\mathcal{E}\). These singularities may correspond to critical configurations where the thermodynamic description breaks down, possibly reflecting extreme fragility or instability of the sector when represented by a single economic agent.

\section{Discussion and Conclusions}

This work proposes an econophysics approach to the sectoral dynamics of Bogotá’s Sports Satellite Account (CSDB), grounded in statistical thermodynamics and geometrothermodynamics. This conceptual framework enables the interpretation of the economy as a complex system in which money plays a role analogous to energy, and thermodynamic quantities—such as entropy, temperature, and heat capacity—acquire well-defined economic meanings.

In particular, sectors \(\mathbb{S}_{15}\) (gambling and betting) and \(\mathbb{S}_{16}\) (recreational and sports activities) provide an illustrative contrast. We observed that \(S_{\mathbb{S}_{15}} < S_{\mathbb{S}_{16}}\), indicating that the betting sector exhibits a higher degree of organization and more efficient information management—largely attributable to stringent government regulation. In contrast, the heterogeneous and decentralized nature of recreational and sports activities results in greater economic disorder.

The analysis further revealed that \(\langle m_{\mathbb{S}_{16}} \rangle > \langle m_{\mathbb{S}_{15}} \rangle\), suggesting that agents in the sports sector require substantially more monetary resources per unit of activity, primarily due to infrastructure costs, specialized inputs, and reliance on imports. Consistently, the heat capacity satisfies \(C_{\mathbb{S}_{16}} > C_{\mathbb{S}_{15}}\), implying that the sports sector demands greater activation energy—or initial investment—to sustain its economic dynamics. Conversely, the betting sector benefits from minimal physical infrastructure requirements, leveraging digital platforms, widespread mobile device penetration, and affordable internet connectivity to operate efficiently at scale.

The directional heat-like flow \(\Delta Q_{\mathbb{S}_{15} \to \mathbb{S}_{16}}\) reflects an implicit resource transfer from the betting sector to the recreational-sports sector—a finding consistent with real-world fiscal mechanisms, such as earmarked taxes on gambling that fund public sports initiatives. This thermodynamic analogy thus offers a quantitative lens to analyze inter-sectoral redistribution policies.

Moreover, the geometrothermodynamic analysis of the equilibrium manifold \(\mathcal{E}\) reveals curvature singularities in the Ricci and Kretschmann scalars, interpreted as  phase transitions  in the economic system—i.e., critical points associated with sectoral crises or structural reconfigurations. These geometric invariants thus emerge as early-warning indicators of instability.

Our results demonstrate that economic entropy serves as an effective measure of sectoral organization: more regulated sectors exhibit lower entropy, suggesting a direct link between institutional oversight and systemic coherence. Similarly, economic heat capacity proves valuable in quantifying the investment intensity required to sustain business activity. The thermodynamic framework not only captures resource flows between sectors but also provides a novel quantitative tool for evaluating redistributive policies. Meanwhile, geometrothermodynamics establishes itself as a promising approach for anticipating critical transitions in economic systems through curvature-based diagnostics.

For future research, we propose extending this analysis to all 17 sectors of the CSDB to map more realistic interaction networks; incorporating longer and higher-frequency time series to validate model robustness; benchmarking this approach against conventional macroeconomic indicators to assess its predictive power during crises or expansions; and enriching the partition function with microeconomic data on agent counts, firm size distributions, and consumption patterns.

In summary, econophysics and statistical thermodynamics offer powerful theoretical frameworks for understanding the sectoral dynamics of Bogotá’s sports economy. Their capacity to translate abstract physical concepts into actionable policy indicators—particularly for foresight and crisis anticipation—merits deeper exploration, especially in the design of evidence-based public interventions.



\end{document}